# SEARCH FOR NEUTRINOLESS DOUBLE BETA DECAY WITH THE NEMO-3 DETECTOR : FIRST RESULTS

D. LALANNE

*Laboratoire de l'Accélérateur Linéaire, IN2P3-CNRS et Université de Paris-Sud*
*91405 Orsay, France*

The NEMO-3 detector, which has been operating in the Fréjus Underground Laboratory since February 2003, is devoted to searching for neutrinoless double beta decay ($\beta\beta0\nu$). The expected performance of the detector has been successfully achieved. Half-lives of the two neutrinos double beta decay ($\beta\beta2\nu$) have been measured for $^{100}$Mo, $^{82}$Se, $^{96}$Zr, $^{116}$Cd and $^{150}$Nd. After 265 days of data collection from February 2003 until March 2004, no evidence for neutrinoless double beta decay ($\beta\beta0\nu$) was found from ~7 kg of $^{100}$Mo and ~1 kg of $^{82}$Se. The corresponding lower limits for the half-lives are $3.5 \times 10^{23}$ years at 90% C.L for $^{100}$Mo and $1.9 \times 10^{23}$ years for $^{82}$Se. Limits for the effective Majorana neutrino mass are $<m_\nu> < 0.7 - 1.2$ eV for $^{100}$Mo and $<m_\nu> < 1.3 - 3.2$ eV for $^{82}$Se. Radon is the dominant background today and a radon trapping factory will be in operation by the end of September 2004. The NEMO-3 expected sensitivity after 5 years of data is 0.2 eV.

## 1. Introduction

Neutrinoless double beta decay ($\beta\beta0\nu$) is a process beyond the Standard Model which violates lepton number by 2 units. The discovery of this decay would be experimental proof that the neutrino is a Majorana particle. It would also constrain the mass spectrum and the absolute mass of the neutrinos. The NEMO-3 detector is devoted to searching for $\beta\beta0\nu$ decay with the direct detection of the two electrons from $\beta\beta$ decay by a combination of a tracking device and a calorimeter.

## 2. The NEMO-3 detector

The NEMO-3 detector [1], installed in the Fréjus Underground Laboratory (LSM, France) is a cylinder divided into 20 equal sectors. The isotopes present inside the detector in the form of very thin foils (30-60 $mg/cm^2$) are $^{100}Mo$ (6914 g), $^{82}Se$ (932 g), $^{116}Cd$ (405 g), $^{130}Te$ (454 g), natural Te (491 g), $^{150}Nd$ (34 g), $^{96}Zr$ (9 g), $^{48}Ca$ (7 g) and Cu (321 g). Natural Te and Cu are devoted to measuring the external background. The sources have been purified to reduce their content of $^{214}$Bi and $^{208}$Tl. On both sides of the sources, there is a gaseous tracking detector. It consists of 6180 open drift cells operating in the Geiger mode regime (Geiger cells) which gives three-dimensional track reconstruction, followed by a calorimeter of 1946 plastic scintillator blocks coupled to very low radioactive photomultipliers (PMTs) for energy measurement. A solenoid surrounding the detector produces a 25 G magnetic field. Finally an external shield of 18 cm of iron and 35 cm of water covers the detector to reduce external $\gamma$ and neutron fluxes. Thus the combination of a tracking detector, a calorimeter and a magnetic field allows the identification of electrons, positrons, $\gamma$ and $\alpha$ particles.

## 3. The performance of the detector

Within the tracking detector, 99.5% of the geiger cells are functioning normally. The resolution on the vertex position of the two reconstructed tracks is $\sigma_t = 0.6$ cm in the transverse plane and $\sigma_l = 1.3$ cm in the longitudinal plane. The ambiguity between $e^-$ and $e^+$ is 3% at 1 MeV.

Within the calorimeter, 97% of the PMTs coupled to scintillators are functioning correctly. The energy resolution using $^{207}Bi$ sources, is 15% (FWHM) at 1 MeV for the 5'' PMTs on the external wall and 17% for the 3'' PMTs on the internal wall. A daily laser survey controls the gain stability of each PMT. The efficiency to detect a $\gamma$ at 500 keV is about 50% with a treshold of 30 keV. The time resolution measured with the 2 electron channel, is 250 ps at 1 MeV which is much smaller than the time-of-flight of a crossing electron that is larger than 3 ns. Thus external crossing electrons are totally rejected.

## 4. Measurement of $\beta\beta 2\nu$ decays with several nuclei

The detector has been running since February 2003. The trigger configuration requires at least 1 PMT with an energy above 150 keV and 3 active geiger cells. The trigger rate is ~7 Hz. A $\beta\beta$ event is an event with 2 tracks coming from the same vertex on the foil. Each track is associated to a fired scintillator with a good internal time-of-flight hypothesis, and the curvature corresponds to a negative charge. After 241 days of data analysed, more than 140,000 $\beta\beta 2\nu$ events from ~7 kg of $^{100}Mo$ have been measured. Figure 1 shows the spectrum of the summed energy of the two electrons for $^{100}Mo$ after background subtraction which is in agreement with the expected spectrum from $\beta\beta 2\nu$ simulation. The signal-to-background ration is 46. The preliminary value of the measurement half-life is $7.72\pm0.02(stat)\pm0.54(syst)10^{18}$y. The $\beta\beta 2\nu$ decay has also been measured for $^{82}$Se, $^{96}$Zr, $^{116}$Cd and $^{150}$Nd. (see table 1).

## 5. Study of the background in the $\beta\beta 0\nu$ energy window

Each background component can be measured using different channels in the data.

External $^{214}$Bi and $^{208}$Tl backgrounds (mostly inside PMTs) have been measured by searching for external ($e^-$, $\gamma$) events in the data. The total activity deduced of $^{208}$Tl is ~40 Bq and is in agreement with previous HPGe measurements of a sample of the PMTs glass. The expected number of $\beta\beta 0\nu$-like events is negligible, $\lesssim 10^{-3}$ counts kg$^{-1}$y$^{-1}$ in the [2.8-3.2] MeV energy window where $\beta\beta 0\nu$ signal is expected.

External neutrons and high energy $\gamma$ backgrounds have been measured by searching for internal ($e^-$, $e^-$) events above 4 MeV. Only 2 events have been observed in 265 days of data collection, as expected in the Monte-Carlo. This background is also negligible, $\lesssim$ 0.02 counts kg$^{-1}$y$^{-1}$ in the window [2.8-3.2] MeV.

The level of $^{208}$Tl impurities inside the molybdenum sources have been measured by searching for internal ($e^-$, $\gamma$) and ($e^-$, $\gamma\gamma$) events. An activity of ~100 µBq/kg has been measured in agreement with HpGe, measurements done before installing the sources in the detector. This corresponds to $\beta\beta 0\nu$-like events of ~0.1 counts kg$^{-1}$y$^{-1}$ in the [2.8-3.2] MeV $\beta\beta 0\nu$ energy window.

The tail of the $\beta\beta 2\nu$ is ~0.3 counts kg$^{-1}$y$^{-1}$ in the [2.8-3.2] MeV $\beta\beta 0\nu$ energy window.

The dominant background today is the radon inside the tracking chamber due to a diffusion of radon from the laboratory (~15 Bq/m$^3$) into the detector. A high efficiency radon detector has measured radon in the NEMO-3 gas. Radon can also be measured directly by searching for ($e^-$, $\alpha$) events in the NEMO-3 data with delayed alphas emitted by $^{214}$Po in the *Bi – Po* process. Both measurements are in good agreement and indicate a level of radon inside the detector of ~20-30 mBq/m$^3$. This radon contamination corresponds to an expected number of $\beta\beta 0\nu$-like events of ~1 count kg$^{-1}$y$^{-1}$ in the [2.8-3.2] MeV $\beta\beta 0\nu$ energy window, a factor ~10 too high to reach the expected sensitivity. A radon trapping factory, designed to reduce radon contamination by a factor ~100 will be in operation by the end of September 2004.

## 6. Preliminary results on the limit of $\beta\beta 0\nu$ decay with $^{100}$Mo and $^{82}$Se

Figures 2 and 3 show the spectrum of the energy sum of the two electrons in the $\beta\beta 0\nu$ energy window for $^{100}$Mo and $^{82}$Se. The number of two electron events observed in the data is in agreement with the expected number of events from $\beta\beta 2\nu$ and the radon simulations. In the energy window [2.8-3.2] MeV, the expected background is 7.0±1.7 and 8 events have been observed from $^{100}$Mo. To check independently the dominant radon contribution above 2.8 MeV, the same energy sum spectrum (Fig. 4) has been plotted for the two electrons emitted from the copper and tellurium foils where no background except radon is expected : the data are in agreement with the radon simulation.

The NEMO-3 detector is able to measure not only the energy sum ($E_{tot}$) of the 2e$^-$ events but also the single energy ($E_{min}$ energy of the e$^-$ of minimum energy) and the angle between the two tracks ($\cos\theta$). Moreover the level of each component of background can be measured. Therefore a maximum likelihood analysis has been applied on 2e$^-$ events above 2 MeV using these three variables [2]. A three-dimensional probability distribution function, $P^{3D}$, can be written as :

$$P^{3D} = P(E_{tot})P(E_{min}/E_{tot})P(\cos\theta/E_{min})$$

where $P(E_{min}/E_{tot})$ and $P(\cos\theta/E_{min})$ are two conditional probability distribution functions. The likelihood is defined as

$$L = \prod_{i=1}^{Ntot}\left(\sum_{k=1}^{8}x_k P_{3D}^k\right)$$

where $k$ corresponds to one of the 8 contributions $\beta\beta 0\nu$, $\beta\beta 2\nu$, radon, external and internal $^{214}$Bi and $^{208}$Tl, and neutrons. The ratio $x_k$ is the number of 2e$^-$ events due to the process $k$ to the total number of observed events $N_{tot}$, and $P_{3D}^K$ is built using simulated events of the contribution $k$. The only free parameter is the ratio $x_{0\nu}$.

With 265 days of data, limits obtained with the likelihood analysis are $T_{1/2}(0\nu) > 3.5 \times 10^{23}$ years at 90% C.L for $^{100}$Mo and $1.9 \times 10^{23}$ years for $^{82}$Se. The corresponding upper limits for the effective Majorana neutrino mass range from 0.7 to 1.2 eV for $^{100}$Mo and 1.3 to 3.6 eV for $^{82}$Se depending on the nuclear matrix elements [3-6]. Limit on Majoron is $T_{1/2}(M) > 1.4 \times 10^{22}$ years at 90% C.L, corresponding to a limit of $\chi < (5.3-8.5) \times 10^{-5}$ [3-4].

## Conclusion

The NEMO-3 detector has been running reliably since February 2003. The $\beta\beta 2\nu$ decay has been measured for $^{82}$Se, $^{96}$Zr, $^{100}$Mo, $^{116}$Cd and $^{150}$Nd. All components of the background in the $\beta\beta 0\nu$ energy window have been measured directly using different channels in the data..The energy sum spectrum of 2e$^-$ events is in agreement with the simulations. After 265 days of data, no evidence for $\beta\beta 0\nu$ decay is found from the 6.914 kg of $^{100}$Mo and 0.932 kg of $^{82}$Se. A likelihood analysis gives an upper limit for the effective neutrino mass $<m_\nu> < 0.7 - 1.2$ eV for $^{100}$Mo. After radon purification and 5 years of data collection, the expected sensitivity will be $T_{1/2}(0\nu) > 4 \times 10^{24}$ years at 90% C.L for $^{100}$Mo and $8 \times 10^{23}$ years for $^{82}$Se, corresponding to $<m_\nu> < 0.2$-$0.35$ eV for $^{100}$Mo and $<m_\nu> < 0.65$-$1.8$ eV for $^{82}$Se.

## References

1. Arnold and al. Physics/0402115, accepted for publication in *Nucl. Inst. Meth. A.*
2. Etienvre, PhD Thesis, University Paris-Sud (2003).
3. Simkovic and al., *Phys. Rev.* **C60** (1999)
4. Stoica and al., *Nucl. Phys.* **A694** (2001)
5. Rodin and al., *Phys. Rev.* **C68** (2003)
6. Caurier and al., *Phys. Rev* **Lett. 77** (1996)

Table 1
Preliminary results of the measurements of $\beta\beta2\nu$ decays

| Isotope | mass (g) | days of data | number of $\beta\beta$ events | Signal/ Background | $T_{1/2}(\beta\beta2\nu)$ (years) |
|---|---|---|---|---|---|
| $^{82}Se$ | 932 | 241.5 | 2385 | 3.3 | $10.3 \pm 0.2(stat) \pm 1.0(syst)10^{19}$ y. |
| $^{96}Zr$ | 9.4 | 168.4 | 72 | 0.9 | $2.0 \pm 0.3(stat) \pm 0.2(syst)10^{19}$ y. |
| $^{100}Mo$ | 6914 | 241.5 | 145245 | 45.8 | $7.72 \pm 0.02(stat) \pm 0.54(syst)10^{18}$ y. |
| $^{116}Cd$ | 405 | 168.4 | 1371 | 7.5 | $2.8 \pm 0.1(stat) \pm 0.3(syst)10^{19}$ y. |
| $^{150}Nd$ | 37.0 | 168.4 | 449 | 2.8 | $9.7 \pm 0.7(stat) \pm 1.0(syst)10^{18}$ y. |

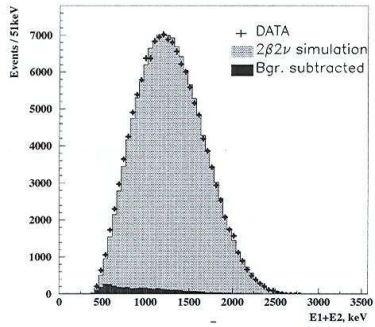

Figure 1. Spectrum of the energy sum of the two electrons from 6.914 kg of $^{100}$Mo after 241 days of data collection.

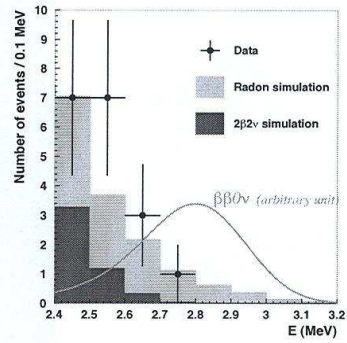

Figure 3. Spectrum of the energy sum of the two electrons above 2.4 MeV from 0.932 kg of $^{82}$Se after 265 days of data collection.

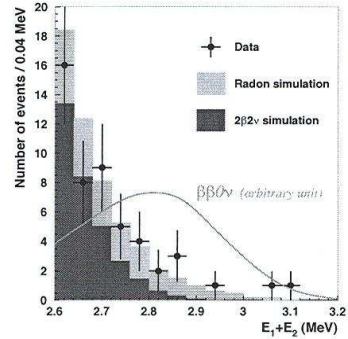

Figure 2. Spectrum of the energy sum of the two electrons above 2.6 MeV from 6.914 kg of $^{100}$Mo after 265 days of data collection.

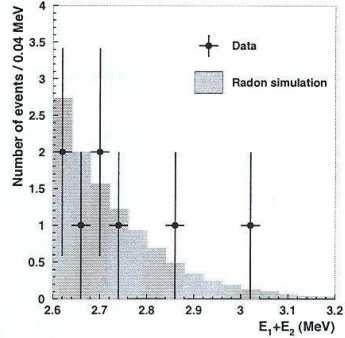

Figure 4. Spectrum of the energy sum of the two electrons above 2.6 MeV from copper and tellerium foils after 265 days of data collection.